\documentclass[12pt,a4paper]{article}
\usepackage{amssymb}
\usepackage{mathrsfs}

\title{\pltp y as canonical \tfn }

\def\ba{\begin{array}}
\def\ea{\end{array}}
\def\be{\begin{equation}}
\def\ee{\end{equation}}
\def\bea{\begin{eqnarray}}
\def\eea{\end{eqnarray}}

\def \rf  {(\ref}
\def \lif  {left--invariant field}

\def\eqn{equation}
\def\tfn{transformation}

\def\cond{condition}

\def\+{{+\!\!\!+}}
\def\-1{^{-1}}
\def\itr{^{-t}}
\def\half{\frac{1}{2}}
\def\unit{{\bf 1}}

\def\el{{e}^{L}}

\def\vl{{v}^{L}}
\def\real{{\mathbb R}}

\def\dxp{{\partial_+}}
\def\dxm{{\partial_-}}
\def\dpm{{\partial_\pm}}
\def\cp{{\cal {P}}}
\def\cF{{\cal {F}}}
\def\cB{{\cal {B}}}
\def\cY{{\cal {Y}}}
\def\hcp{\widehat \cp}
\def\ls{L_\sigma}
\def\hls{\widehat L_\sigma}
\def\ch{\mathscr{H}}
\def\lagr{\mathscr{L}}

\def\sm{$\sigma$--model}
\def\pl{Poisson--Lie }
\def\pltp{Poisson--Lie T--pluralit}
\def\pltd{Poisson--Lie T--dualit}
\def\dd{Drinfel'd double}

\def\wt{\widehat}
\def\dotplus{\stackrel{{\bf.}}{+}}

\def\rd{{\rm d}}

\def\ce{{\cal {E}}}
\def\cJ{{\cal {J}}}
\def\cd{{\mathfrak d}}

\def\cg{{\mathfrak g}}
\def\tcg{{\tilde {\mathfrak g}}}

\def\ttil{\tilde{T}}
\def\htil{\tilde{h}}

\def\rpg{L_-(g)}
\def\rmg{L_+(g)}
\def\rphatg{\wt L_-(\hat g)}
\def\rmhatg{\wt L_+(\hat g)}
\def\rpmtilh{\tilde L_\pm(\tilde{h})}
\def\rptilh{\tilde L_-(\tilde{h})}
\def\rmtilh{\tilde L_+(\tilde{h})}
\def\that{\widehat{T}}
\def\ghat{\hat{g}}

\author{Ladislav Hlavat\'y and Libor \v Snobl
\\ {\it Department of Physics,}
\\ {\it Faculty of Nuclear Sciences and Physical Engineering,}
\\{ \it Czech Technical University,}
\\ {\it B\v rehov\'a 7, 115 19 Prague 1, Czech Republic}
\\ {\it {E--mail: Ladislav.Hlavaty@fjfi.cvut.cz}, {Libor.Snobl@fjfi.cvut.cz}}
\\ {\it Phone: +420 224 358 294, Fax: +420 222 320 861
 }}
\begin{document}
\maketitle
\abstract{We generalize the prescription realizing classical Poisson--Lie T--duality as canonical transformation to
Poisson--Lie T--plurality. The key ingredient is the transformation of left--invariant fields under
Poisson--Lie T--plurality. Explicit formulae realizing canonical transformation are presented and the preservation
of canonical Poisson brackets and Hamiltonian density is shown.}
\bigskip

\bigskip


\section{Introduction}

Poisson--Lie T--duality and T--plurality is already quite an old subject. It was introduced in 1995 when C.
Klim\v c\'{\i}k and P. \v Severa in \cite{klse:dna,kli:pltd,klse:pltdlgdd} proposed Poisson--Lie T--duality  as
an approach solving certain problems in T--duality with respect to non--Abelian groups of isometries (especially
that the original T--duality worked only in one direction). Already in \cite{klse:dna,klse:pltdlgdd} they
considered the possibility of what is now called Poisson--Lie T--plurality. This is related to the fact that
the Lie algebra of Drinfel'd double may be decomposable into more than one pair of subalgebras
whose transposition corresponds to duality. On the other hand, further development (like the explicit
formulation of canonical transformation in \cite{sfetsos:pltdss,sfetsos:cenismpltd}) focused only on
Poisson--Lie T--duality and almost no explicit formulae and no examples of genuine Poisson--Lie T--plurality
were known until 2002 when R. von Unge considered T--plurality of conformal
quantum sigma models (on one--loop level) in \cite{unge:pltp} and coined the current phrase ``Poisson--Lie T--plurality''. By
that time classifications of Drinfel'd doubles in low dimension, e.g. \cite{snohla:ddoubles}, became available,
facilitating construction of more examples and study of their properties (see \cite{hlasno:3dsm2} and references
therein).

Gradually, the need arose for generalization of formulae previously derived in the Poisson--Lie T--duality context
to the general plurality case. One of these is the formulation of Poisson--Lie T--plurality as canonical
transformation (derived for the duality case by K. Sfetsos in \cite{sfetsos:pltdss,sfetsos:cenismpltd}).

In this paper we shall derive the explicit canonical formulation of \pltp y. As we shall show, the key point is
the \tfn{} of the extremal left and right--invariant fields, which can be derived in a direct way, and which will enable
us to find the \tfn{} of the canonical variables of the dualizable \sm s and prove that they really constitute a
canonical transformation.

One of possible applications of our results is in the study of the worldsheet boundary conditions. Recently, Poisson--Lie
T--duality \tfn{} of worldsheet boundary \cond s of the dualizable \sm s was derived in \cite{are:wsbc}. The key
formulae there were the \tfn s of left--invariant fields by the \pltd y obtained from canonical formulation of
T--dual \sm s \cite{sfetsos:pltdss,sfetsos:cenismpltd}. Using the formulae derived in this paper one can easily
generalize the results of \cite{are:wsbc} to the T--plurality case. Detailed discussion of them shall be
the subject of future work.
\medskip

A note concerning the conventions: we are using in the current paper the conventions introduced in
\cite{sfetsos:pltdss,sfetsos:cenismpltd} in order to be able to compare with results therein. Unfortunately,
this notation is not the same as the one used in \cite{klse:dna,kli:pltd} and all our previous papers. The two
notations are equivalent upon substitutions $g,l, \ldots \leftrightarrow g^{-1}, l^{-1}, \ldots$ accompanied by
the worldsheet parity transformation $x_+ \leftrightarrow x_- $.

\section{Elements of \pltp y}
For simplicity we shall consider the \sm s without spectator fields, i.e. with target manifold isomorphic to the
group of generalized isometries. The inclusion of spectators is straightforward, see \cite{sfetsos:pltdss,unge:pltp}. The classical action
of \sm{} without spectators reads \be S_{\ce}[\phi]=\frac{1}{2} \int d^2x\,
\partial_+ \phi^{\mu}\ce_{\mu\nu}(\phi)
\partial_- \phi^\nu \label{sigm1} \ee where $\ce$ is a tensor on a Lie group $G$ and the functions $ \phi^\mu:\ V\subset\real^2\rightarrow \real,\
\mu=1,2,\ldots,{\dim}\,G$ are obtained by the composition $\phi^\mu=y^\mu\circ g $ of a map
 $g:V\subset\real^2\rightarrow G$ and a coordinate map $y$ of a neighborhood $U_g$ of an element $g(x_+,x_-)\in G$.
Further on we shall use formulation of the \sm s in terms of left--invariant fields $g\-1\dpm g$. The
tensor $\ce$ can be written 
as \be \ce_{\mu\nu}={\el_\mu}^a (g)E_{ab}(g) {\el_\nu}^b(g)\label{metorze} \ee where

\begin{itemize}
\item ${\el_\mu}^a$ are components of left--invariant forms (vielbeins) $g^{-1}{\rm d}g =\rd y^\mu
{\el_\mu}^a(g)T_a$,
\item $T_a$ are basis elements of $\cg$, i.e. Lie algebra of $G$,
\item $E_{ab}(g)$ are matrix elements of a $G$--dependent bilinear nondegenerate form on $\cg$  in the basis $\{T_a\}$.
\end{itemize} The
action of the \sm{} then reads\footnote{Central dot means matrix multiplication and we consider $L_\pm$ as a row
vector whereas $T$ is a column vector with components $T_a$. $L^t$ denotes transposition. Later on we shall also
use the notation  $M^{-t}=(M^{-1})^t$ for matrices.} \be \label{SFg}S[g]=\half\int d^2x\, L_+(g)\cdot E(g)\cdot L_-^t(g) \ee
where \be\label{rpm} L_\pm(g)^a\equiv(g\-1\dpm g)^a=\dpm \phi^\mu {\el_\mu}^a(g),\ \ \ g\-1\dpm g= L_\pm(g)\cdot
T. \ee

The \sm s that can be transformed by the \pltp y are formulated (see \cite{klse:dna,kli:pltd}) on a \dd{}
$D\equiv(G|\tilde G)$ -- a Lie group whose Lie algebra $\cd$ admits a decomposition $\cd=\cg\dotplus\tcg$
into a pair of subalgebras maximally isotropic with respect to a symmetric ad-invariant nondegenerate
bilinear form $\langle\, .\,,.\,\rangle $. The matrices $E(g)$ for such \sm s are of the form \be
E(g)=(E_0^{-1}+\Pi(g))^{-1}, \ \ \ \Pi(g)=b^t(g)\cdot a(g) = -\Pi(g)^t,\label{Fg}\ee where $E_0$ is a
constant matrix and $a(g),b(g)$ are submatrices of the adjoint representation of the subgroup $G$ on the Lie
algebra $\cd$ satisfying\be\label{adgt} gTg\-1\equiv Ad(g)\triangleright T=a\-1(g)\cdot T,\ \ \ \ g\tilde
Tg\-1\equiv Ad(g)\triangleright \tilde T =b^t(g)\cdot T+ a^t(g)\cdot \tilde T, \ee and $\ttil^a$ are
elements of dual basis in the dual algebra $\tcg$, i.e. $\langle\,T_a ,\,\ttil^b\,\rangle=\delta_a^b $. The
matrix $a(g)$ also relates the left-- and right--invariant fields on G
\begin{equation}\label{lra} \dpm gg\-1={R_\pm}(g)\cdot T,\ \ \
    {L_\pm}(g)=R_\pm(g) \cdot a(g).
\end{equation}

The \eqn s of motion of the dualizable \sm s can be written as Bianchi identities for the left--invariant
fields $\rpmtilh$ on the dual algebra $\tcg$
\begin{eqnarray}
 \label{kl10} \rmtilh\cdot
\ttil\equiv\htil\-1\dxp\htil & = &\rmg\cdot E(g)\cdot a^t(g)\cdot \ttil, \\
\nonumber \rptilh\cdot \ttil\equiv\htil\-1\dxm\htil & = & -\rpg\cdot E^t(g)\cdot a^t(g)\cdot \ttil .\end{eqnarray}
This is a consequence of the fact
that the \eqn s of motion of the dualizable \sm{} can be formulated as the \eqn s on the \dd{} \cite{klse:dna}
\begin{equation}\label{ddeqm}
    \langle\, l\-1\dpm l\,,{\cal E^\mp}\,\rangle =0,
\end{equation}
where $l=\htil g\in D, \ \htil\in \tilde G, \ g\in G$ and
$${\cal E^+}={\rm span}\left(T+E_0 \cdot \tilde T \right), \qquad {\cal E^-}={\rm span}\left(T- E_0^t \cdot \tilde T \right)   $$
are two orthogonal subspaces in $\cd$. (The unique decomposition $l=\htil g$ on $D$ exists for a general \dd{} only in the vicinity of the group unit. For the so--called perfect \dd s it is defined globally and we shall consider only these. Otherwise all the constructions considered would hold only locally.)

In general there are several decompositions (Manin triples) of a \dd. Let $\hat\cg\dotplus\bar\cg$ be another
decomposition of the Lie algebra $\cd$ into maximal isotropic subalgebras. Then another \sm{} can be defined.
The dual bases of $\cg,\tcg$ and $\hat\cg,\bar\cg$ are related by the linear \tfn\begin{equation}\label{pqrs}
    \left(\matrix{T \cr\tilde T\cr} \right)= \left(\matrix{K&Q \cr R&S \cr} \right) \left(\matrix{\hat T\cr
\bar T\cr} \right), \end{equation}
where the matrices $K,\, Q,\, R,\, S$ are chosen in such a way that the structure of the
Lie algebra $\cd$
\begin{eqnarray}\label{liestruc}  [T_a,T_b] &=& {f_{ab}}^c T_c,\nonumber\\ {}
[\tilde T^a,\tilde T^b] &=& {{\tilde f}^{ab}}{}_c\tilde T^c,\\ {} [\tilde T^a,T_b] &=&   {f_{bc}}^a \tilde T^c -
{{\tilde f}^{ac}}{}_b T_c\nonumber
\end{eqnarray}transforms to the similar one where $T\rightarrow\that,\ \ttil\rightarrow\bar T$ and
the structure constants $f,\, \tilde f$ of $\cg$ and $\tcg$ are replaced by the structure constants $\hat
f,\,\bar f$ of $\hat \cg$ and $\bar \cg$. The duality of both bases requires
\be\label{trinv} \left(\matrix{K&Q \cr R&S \cr}
\right)\-1=\left(\matrix{S^t&Q^t \cr R^t&K^t \cr} \right).\ee
The other \sm{} is defined analogously
to (\ref{SFg}-\ref{Fg}) where \be \wt E(\ghat)=(\wt E_0^{-1}+\wt\Pi(\ghat))^{-1}, \ \ \ \wt\Pi(\ghat)=\wt
b^t(\ghat)\cdot \wt a(\ghat) = -\wt\Pi(\ghat)^t,\label{Fghat}\ee\be \label{E0hat} \widehat E_0=(K+E_0\cdot
R)\-1\cdot (Q+E_0\cdot S)=(S^t\cdot E_0-Q^t)\cdot (K^t-R^t\cdot E_0)\-1,\ee and classical solutions of the two \sm s are
related by two possible decompositions of $l\in D$,
\begin{equation}\label{lgh}
    l=\htil g=\bar h \ghat.
\end{equation}
The explicit examples of solutions of the \sm s related by the \pltp y are given in \cite{hlahytur}.

\section{\pl \tfn{} of extremal \lif s}
As mentioned in the Introduction, the formulae for \tfn{} of \lif s evaluated on solutions of \eqn s of motion
(hence extremal) by the \pltd y were found in \cite{are:wsbc}.
We are going to derive the extension of these formulae in an
alternative way.

Let us write the \lif{} $l\-1\dxp l$ on the \dd{} in terms of $\rmg$ and
$\rmtilh$
\begin{eqnarray}\label{rgthil}
    l\-1\dxp l &=& (\htil g)\-1(\dxp (\htil g))=\rmg\cdot T+\rmtilh\cdot g\-1\ttil
    g\nonumber\\ &=&\rmg\cdot T+\rmtilh\cdot \left[b(g)\cdot T+a\itr(g)\cdot \ttil\right]
\end{eqnarray}
where $a(g)$ and $b(g)$ are the matrices introduced in (\ref{adgt}).

Using the \eqn s of motion (\ref{kl10}) and the expression (\ref{Fg}) for $E(g)$ we get\bea\label{rghtil2}
    l\-1\dxp l=  &=&\rmg\cdot T+\rmg\cdot E(g)\cdot \left[a^t(g)\cdot b(g)\cdot T+\ttil\right]\nonumber\\
    &=&\rmg\cdot E(g)\cdot \left[E_0\-1\cdot T+\ttil\right].
\eea

On the other hand, from the decomposition $l=\bar h \ghat$ we find by a similar procedure
\be\label{rghathp}
    l\-1\dxp l=
   \rmhatg\cdot \wt E(\hat g)\cdot \left[\wt E_0\-1\cdot \that+\bar T\right].
\ee Inserting (\ref{pqrs}) and (\ref{E0hat}) into (\ref{rghtil2}) and comparing coefficients of $\that$ and $\bar T$ with those in
\rf{rghathp}) we obtain the formula for \tfn{} of the \lif s under the \pltp y \be \label{rphatg} \rmhatg=\rmg\cdot
E(g)\cdot \left[S+E_0\-1\cdot Q\right]\cdot {\wt E}\-1(\hat g). \ee In the same way we can derive \be
\label{rmhatg} \rphatg =\rpg\cdot E^t(g)\cdot \left[S-E_0\itr\cdot Q\right]\cdot \wt E\itr(\hat g).\ee

This agrees with the formulae obtained in \cite{are:wsbc} for \pltd y, i.e. for $Q=R=\unit,\ K=S=0,$ $\tilde
L_{\pm}(\tilde g)=\wt L_{\pm}(\ghat)$, which in our notation (i.e. $L_\pm$ rows) read
\begin{eqnarray} \label{are37}
\tilde L_{+}^t(\tilde g) & = & \tilde E\itr(\tilde g) \cdot E_0\itr  \cdot E^t(g) \cdot L_+^t(g), \\
\nonumber \tilde L_{-}^t(\tilde g) & = & -(\tilde E(\tilde g))\-1 \cdot E_0\-1  \cdot E(g) \cdot L_-^t(g).
\end{eqnarray} The \tfn s of right--invariant extremal fields can be easily obtained from the relation (\ref{lra}).

\section{Transformation of canonical variables}

In the present section we are going to generalize the formulae for canonical transformation obtained in \cite{sfetsos:pltdss,sfetsos:cenismpltd} for the Poisson--Lie T--duality to the general T--plurality case.

Recall that the time and space coordinates on the worldsheet are $\tau=x_+ + x_-, \; \sigma =x_+ -x_-$, i.e.
$\partial_\tau=\frac{1}{2} (\partial_+ +\partial_-), \; \partial_\sigma=\frac{1}{2} (\partial_+
-\partial_-)$. The canonical momentum is defined by \be \cp_\mu = \frac{\partial \lagr}{\partial
(\partial_\tau \phi^\mu)} = \frac{1}{2} \left( \ce_{\mu \nu} \partial_- \phi^\nu +  \ce_{\nu \mu} \partial_+
\phi^\nu  \right).\label{momentum_a}\ee It turns out, similarly as above, to be advantageous to use  a
momentum in local frame, defined as \be \cp_a = {\vl_a}^{\mu}(g) \cp_\mu \ee where $\vl = (\el)^{-1}$. We
shall denote by $\cp$ the column vector with the components $\cp_a$ so that \be\label{momentum_a2}
\cp=\frac{1}{2} \left( E(g) \cdot  L_-^t(g)+ E^t(g) \cdot  L_+^t(g)\right). \ee We also define \be \ls =
\frac{1}{2} \left( L_+(g)-L_-(g) \right) .\ee For the future reference let us quote the inverse relations
\begin{eqnarray}\label{invpls} L_+(g) &=& 2 \left(\cp^t+\ls \cdot E^t(g) \right) \cdot  \left( E(g)+E^t(g)
\right)^{-1}, \\ \nonumber L_-(g) &=& 2 \left( \cp^t-\ls \cdot E(g) \right) \cdot \left( E(g)+E^t(g)
\right)^{-1}. \end{eqnarray} Defining the similar quantities $\hcp, \hls$ for the model after the
Poisson--Lie T--plurality transformation and using (\ref{rphatg},\ref{rmhatg}) we find \be\label{canplurpA}
 \hcp= \frac{1}{2} \left( ( Q^t  \cdot E_0^{-t}+S^t) \cdot  E^t(g) \cdot  L_+^t(g)   - (Q^t  \cdot E_0^{-1}-S^t) \cdot
 E(g)  \cdot L_-^t(g)   \right),  \ee
\be\label{canplurlsA}
\hls= \frac{1}{2} \left( L_+(g) \cdot  E(g) \cdot  (E_0^{-1} \cdot  Q+S) \cdot  \widehat E(\hat g)^{-1}
+L_-(g)  \cdot E^t(g) \cdot (E_0^{-t} \cdot Q-S) \cdot  \widehat E(\hat g)^{-t} \right),  \ee
which, as we shall show, becomes the transformation of the canonical variables\footnote{We slightly
abuse the terminology here: strictly speaking the canonical variables are $\cp_\mu,\phi^\mu$ and
$\hat\cp_\mu,\hat\phi^\mu$, respectively. Because the plurality transformation of $\phi^\mu$ defined via
(\ref{lgh}) (where $h$,$\tilde h$ are constructed via (\ref{kl10})) is nonlocal, we write instead the
transformation of its space derivative $\partial_\sigma \phi^\mu$ and also we use for convenience the local frame versions
instead of coordinate versions of these. Nevertheless, as we show later on, this doesn't lead to any
nonlocalities in the Hamiltonian or the Poisson brackets.}
\begin{eqnarray}  \hcp & = & \left( Q^t \cdot \Pi(g) +S^t \right) \cdot  \cp + Q^t  \cdot \ls^t,\label{canplurp} \\
\label{canplurls} \hls & = & \cp^t  \cdot \left[ \left(S-\Pi(g) \cdot  Q  \right) \cdot  \widehat\Pi(\ghat)+
R-\Pi(g)  \cdot K \right]+ \ls \cdot  \left( Q \cdot  \widehat\Pi(\ghat)+K \right).
\end{eqnarray}
This agrees with the formulae obtained in \cite{sfetsos:pltdss} for \pltd y\footnote
{and, as another consistency check, reduces to identity transformation when $K=S=\unit,R=Q=0$.},
i.e. for $Q=R=\unit,\ K=S=0$, but generalizes the results from \cite{sfetsos:pltdss,sfetsos:cenismpltd} to any T--plurality transformation.
\medskip

In order to deduce (\ref{canplurp},\ref{canplurls}) we shall first list a few useful formulae. Because of their complexity we shall suppress the $g, \tilde g$--dependence in the proof (i.e. till the end of this section) and also the explicit dot $\cdot$ for matrix multiplication. This doesn't lead to any difficulty because the derivation of (\ref{canplurp},\ref{canplurls}) from (\ref{canplurpA},\ref{canplurlsA}) is purely algebraic.

We have matrix identities valid for any matrix $A$ (whenever the expressions make sense)
\begin{eqnarray}
\nonumber A^{-1} (A^{-1}+A^{-t})^{-1} & = & (A+A^t)^{-1} A^{t}, \\
\label{beta} A^{-t} (A^{-1}+A^{-t})^{-1} & = & (A+A^t)^{-1} A, \\
\nonumber A^{-1} (A^{-1}+A^{-t})^{-1} A^{-t} & = & (A+A^t)^{-1} =  A^{-t} (A^{-1}+A^{-t})^{-1} A^{-1}.
\end{eqnarray}
Directly from the definition (\ref{Fg}) of $E$ we have
\be\label{eee0e0}
E^{-1}+E^{-t}=E_0^{-1}+E_0^{-t}
\ee
and its consequences due to (\ref{beta})
\begin{eqnarray}
\nonumber E(E+E^t)^{-1}E^t & = & (E^{-1}+E^{-t})^{-1}=(E_0^{-1}+E_0^{-t})^{-1},\\
\label{alpha} E^t(E+E^t)^{-1}E & = & (E^{-1}+E^{-t})^{-1}=(E_0^{-1}+E_0^{-t})^{-1}.
\end{eqnarray}
Finally, combining (\ref{alpha}) and (\ref{beta}) (using first $A=E$ and then $A=E_0^{-1}$) together with (\ref{Fg}) (in the second equality) we get
\begin{eqnarray}
\nonumber (E+E^t)^{-1} \left( E E_0^{-1}-E^t E_0^{-t} \right) & = & E^{-t} (E_0^{-1}+E_0^{-t})^{-1}  E_0^{-1} - E^{-1} (E_0^{-1}+E_0^{-t})^{-1}  E_0^{-t} = \\
\nonumber & = & E_0^{-t} (E_0^{-1}+E_0^{-t})^{-1}  E_0^{-1} - E_0^{-1} (E_0^{-1}+E_0^{-t})^{-1}  E_0^{-t}  \\
\nonumber  & & - \Pi (E_0^{-1}+E_0^{-t})^{-1} (E_0^{-1}+E_0^{-t}) = \\
\label{gamma} & = & -\Pi
\end{eqnarray}
and similarly its transpose
\be\label{gamma2}
 \left( E_0^{-t} E^t -  E_0^{-1} E  \right) (E+E^t)^{-1} = \Pi.
\ee

Now, when we substitute the relations (\ref{invpls}) into the formula (\ref{canplurpA}), we get
\begin{eqnarray*}
\hcp & = & \left[ (Q^t E_0^{-t} +S^t) E^t (E+E^t)^{-1} - (Q^t E_0^{-1} -S^t) E (E+E^t)^{-1}\right] \cp + \\
& + &  Q^t \left[E_0^{-t} E^t (E+E^t)^{-1} E + E_0^{-1} E (E+E^t)^{-1} E^t \right] \ls^t
\end{eqnarray*}
(the terms involving $S^t(\ldots)\ls^t$ cancel each other). Using (\ref{alpha}) we simplify the coefficient of $\ls^t$, getting the desired $Q^t \ls^t$ term in (\ref{canplurp}). The terms of the form $S^t(\ldots)\cp$ give $S^t \cp. $
The remaining $Q^t(\ldots) \cp$ terms are simplified using (\ref{gamma2})
$$ Q^t\left( E_0^{-t} E^t - E_0^{-1} E  \right) (E+E^t)^{-1} \cp = Q^t \Pi \cp.$$
Therefore, the formula (\ref{canplurp}) is proven.

Similarly, we substitute the relations (\ref{invpls}) together with the definition of $\widehat E$ (\ref{Fghat},\ref{E0hat}), i.e.
$$ \widehat  E = \left( (Q+E_0 S)^{-1} (K+E_0 R)  + \widehat\Pi \right)^{-1} = \left( (E_0^t S-Q)^{-1} (K-E_0^t R)- \widehat\Pi \right)^{-t}$$
into the formula (\ref{canplurlsA}). We get
\begin{eqnarray*}
\hls & = & \cp^t (E+E^t)^{-1} \left[ E(E_0^{-1} Q+S) \left((Q+E_0 S)^{-1} (K+E_0 R)+ \widehat\Pi \right)\right. +\\
& & + \left. E^t (E_0^{-t} Q-S) \left((E_0^t S-Q)^{-1} (K-E_0^t R)- \widehat\Pi \right)  \right]+ \\
& & + \ls \left[ E^t (E+E^t)^{-1}E (E_0^{-1} Q+S) \left((Q+E_0 S)^{-1} (K+E_0 R)+ \widehat\Pi \right) \right. -\\
& &  \left. -E(E+E^t)^{-1}E^t (E_0^{-t} Q-S) \left((E_0^t S-Q)^{-1} (K-E_0^t R)- \widehat\Pi \right) \right].
\end{eqnarray*}
We note that
$$(E_0^{-1} Q+S) (Q+E_0 S)^{-1} = E_0^{-1}, \qquad (E_0^{-t} Q-S) (E_0^t S-Q)^{-1} = -E_0^{-t}$$
and using relations (\ref{beta}--\ref{gamma}) we simplify the expression for $\hls$ to the desired form (\ref{canplurls}) which finishes the proof of the formulae (\ref{canplurp},\ref{canplurls}).

\section{Poisson--Lie T--plurality as canonical transformation}
In order to show that (\ref{canplurp},\ref{canplurls}) is really a canonical transformation we shall use the
expressions for Poisson brackets of $\cp_a$ and
\be \cJ^a = \ls^a+\Pi(g)^{ab}\cp_b, \ \ {\rm i.e.} \ \cJ=\ls^t+\Pi(g)\cdot \cp\label{J}\ee
introduced in \cite{sfetsos:cenismpltd}, namely
\begin{eqnarray}\label{pbjp}
    \{\cJ^a,\cJ^b\} &=& {{\tilde f}^{ab}}{}_c \cJ^c \delta(\sigma-\sigma'),\nonumber\\
    \{\cp_a,\cp_b\} &=& { f_{ab}}^c \cp_c \delta(\sigma-\sigma'),\\
    \{\cJ^a,\cp_b\} &=& ( {f_{bc}}^a \cJ^c - {{\tilde f}^{ac}}{}_b \cp_c)\delta(\sigma-\sigma') + \delta_b^a\delta'(\sigma-\sigma').\nonumber
\end{eqnarray}
These Poisson brackets are equivalent to the canonical ones
\begin{eqnarray}\label{canpb}
    \{\cp_\mu,\cp_\nu\} &=& \{\partial_\sigma \phi^\mu,\partial_\sigma \phi^\nu\} = 0,\nonumber \\
    \{ \partial_\sigma \phi^\mu, \cp_\nu \} &=&  \delta_\nu^\mu \delta'(\sigma-\sigma').
\end{eqnarray}

Further, using the definition (\ref{J}) we note that the \tfn{} of the canonical momentum (\ref{canplurp}) can be written as \be \hcp = S^t \cdot  \cp + Q^t  \cdot \cJ \label{phat}\ee
which reminds of the inverse of the \tfn{} (\ref{pqrs}) of the basis elements of the \dd. From this one can
conjecture that \be \label{canplurj} \wt \cJ= R^t \cdot \cp+K^t \cdot  \cJ. \ee and a simple calculation using the
definition (\ref{J}) of $\cJ$ proves that (\ref{canplurj}) is indeed equivalent to (\ref{canplurls}).

To prove the invariance of the Poisson brackets \rf{pbjp}) (and thus of (\ref{canpb})) under the \pltp y
\tfn s (\ref{canplurp},\ref{canplurls}) or (\ref{phat},\ref{canplurj}) it is useful to note that their
structure strongly reminds of the Lie structure (\ref{liestruc}) of the \dd, namely that the Poisson
brackets (\ref{pbjp}) can be written in the compact form \be \label{pby} \{\cY_\alpha,\cY_\beta\} = {
\cF_{\alpha\beta}}^\gamma \cY_\gamma \delta(\sigma-\sigma')+\cB_{\alpha\beta}\delta'(\sigma-\sigma')\ee
where $\alpha,\beta,\gamma=1,\ldots, \dim\ \cd$,
\begin{equation}\label{ypq}
    \cY=\left(\begin{array}{cc}
               \cp \\ \cJ
             \end{array}\right),
\end{equation}${\cF_{\alpha\beta}}^\gamma $ are structure coefficients of the \dd{} and $\cB_{\alpha\beta} $
are matrix elements of the bilinear form $\langle\,.\,,.\,\rangle$ in the basis $T_a,\tilde T^a$ of $\cd$.
From this compact form it is clear that the Poisson brackets (\ref{pby}) are form--invariant under the
transformation (\ref{phat},\ref{canplurj}) that is an analog of the \tfn{} (\ref{pqrs}) of bases of the
\dd{} which transforms $f,\,\tilde f$ to $\hat f,\,\bar f$ and preserves the duality of bases, i.e.
$\cB_{\alpha\beta}$. Consequently, the canonical Poisson brackets are invariant, i.e. (\ref{canpb}) is
transformed by \pltp y to
\begin{eqnarray}
    \{\wt \cp_\mu,\wt \cp_\nu\} &=& \{\partial_\sigma \wt \phi^\mu,\partial_\sigma \wt \phi^\nu\} = 0,\nonumber \\
    \{ \partial_\sigma \wt \phi^\mu, \wt \cp_\nu \} &=&  \delta_\nu^\mu \delta'(\sigma-\sigma').
\end{eqnarray}

Finally, we compute the Hamiltonian density \be\label{ham_def} \ch = \partial_\tau \phi^\mu \cp_\mu - \lagr\ee
where the Lagrangian  density is deduced from the action (\ref{SFg})
$$\lagr=\half L_+(g)\cdot E(g)\cdot L_-^t(g)= \frac{1}{4} \left( L_+(g)\cdot E(g)\cdot L_-^t(g)+L_-(g)\cdot E^t(g)\cdot L_+^t(g)  \right)$$
and we have used an obvious identity valid for any column vector $x$ and matrix $A$
\begin{equation}\label{xAx}
x^t A x = x^t A^t x = \frac{1}{2} x^t (A+A^t) x.
\end{equation}
We recall that due to the definition (\ref{momentum_a},\ref{momentum_a2}) of the canonical momentum  we have
$$\partial_\tau \phi^\mu \cp_\mu = \frac{1}{2} \left(\partial_+ \phi^\mu+\partial_- \phi^\mu\right) \cp_\mu =
\frac{1}{4} \left(L_+(g)+L_-(g)\right) \cdot \left( E(g) \cdot  L_-^t(g)+ E^t(g) \cdot  L_+^t(g)\right)$$
Substituting into the definition of Hamiltonian density (\ref{ham_def}) we find
\be\label{hamilt} \ch=
\frac{1}{4} \left( L_-(g) \cdot  E(g) \cdot  L_-^t(g) + L_+(g)  \cdot E(g) \cdot  L_+^t(g)  \right) \ee where
the substitution for the left--invariant fields $L_-(g),L_+(g)$ in terms of the canonical variables (\ref{invpls}) is understood. 
Performing explicitly the substitution (\ref{invpls}) we get for the Hamiltonian density the formula
\be\label{sfetsham} \ch = \frac{1}{2} \left(\cp^t - \ls  \cdot B \right)  \cdot G^{-1} \cdot \left( \cp+B  \cdot
\ls^t \right)+\frac{1}{2} \ls \cdot  G  \cdot \ls^t \ee used in \cite{sfetsos:pltdss,sfetsos:cenismpltd} where
$G,B$ are symmetric and antisymmetric part of $E(g)$, respectively, i.e.
$$G=\frac{1}{2}(E(g)+E^t(g)), \; B=\frac{1}{2}(E(g)-E^t(g)).$$
The Hamiltonian density of the $\sigma$--model obtained by T--plurality transformation can be written analogously as
$$ \widehat \ch = \frac{1}{4} \left(
\wt L_-(\ghat)  \cdot \wt E(\ghat)  \cdot \wt L_-^t(\ghat) + \wt L_+(\ghat) \cdot  \wt E(\ghat) \cdot  \wt
L_+^t(\ghat) \right) $$
where we assume, as above, that the left--invariant fields $\wt L_-(\ghat),\wt L_+(\ghat)$ are expressed in terms of the new canonical variables. Using the transformation of the left--invariant fields (\ref{rphatg},\ref{rmhatg}) we find
\begin{eqnarray*}
\widehat \ch & = & \frac{1}{4} \left(  L_-(g) \cdot E^t(g) \cdot (S-E_0^{-t}Q) \cdot \wt E(\ghat)^{-t} \cdot (S^t-Q^t E_0^{-1})  \cdot  E(g) \cdot L_-^t(g) \right.+ \\
& & + \left. L_+(g) \cdot E(g) \cdot (S+E_0^{-1}Q) \cdot \wt E(\ghat)^{-t} \cdot (S^t+Q^t E_0^{-t})  \cdot  E^t(g) \cdot L_+^t(g) \right) .
\end{eqnarray*}
Due to the identity (\ref{xAx}) we can replace $\wt E^{-t}(\ghat)$ by $\wt E^{-t}(\ghat)+\wt E^{-1}(\ghat)=\wt E_0^{-t}+\wt E_0^{-1}$.
From the definition  of $\wt E_0$ (\ref{E0hat}) and the duality of bases (\ref{trinv}), i.e.
$$ QR^t= \unit-KS^t , \quad RS^t=-SR^t, \quad KQ^t=-QK^t, \quad RQ^t= \unit -SK^t $$
we get
\begin{eqnarray*}
\widehat \ch & = & \frac{1}{8} \left(  L_-(g) \cdot E^t(g) \cdot (E_0^{-1}+E_0^{-t}) \cdot E(g) \cdot
L_-^t(g)\right. \\ & & \left. + L_+(g) \cdot E(g) \cdot (E_0^{-1}+E_0^{-t})  \cdot  E^t(g) \cdot L_+^t(g)
\right).
\end{eqnarray*}
Using the relation (\ref{eee0e0}) we replace $E_0^{-1}+E_0^{-t}$ by $E^{-1}(g)+E^{-t}(g)$ and employ once again the identity (\ref{xAx}), getting the final result
$$\widehat \ch = \frac{1}{4} \left( L_-(g) \cdot  E(g) \cdot  L_-^t(g) + L_+(g)  \cdot E(g) \cdot  L_+^t(g)  \right).$$
Consequently, we find that the Hamiltonian density is
preserved under Poisson--Lie T--plurality transformation, \be \widehat \ch = \ch. \ee We could have
equivalently used the form of the Hamiltonian density (\ref{sfetsham}) together with the transformation of
canonical variables (\ref{canplurp},\ref{canplurls}). In the approach we used the computation of $\widehat
\ch$ in terms of original canonical variables $\cp,L_\sigma$, or equivalently the left--invariant fields $L_-(g),L_+(g)$, is significantly algebraically simpler.

\section{Conclusions}
We have derived a \tfn{} of the canonical structure of dualizable \sm s,
more precisely their (pseudo)canonical variables, Poisson brackets and Hamiltonian densities
under the \pltp y. It turned out that by a suitable choice of the variables 
the Poisson brackets acquire a rather symmetric form that can be turned into the compact form (\ref{pby}).
This expression is explicitly form--invariant with respect to the choice of basis in the \dd{} on which the \sm s are defined. This proves the invariance of the canonical structure under the \pltp y because its transformations follow from various
decompositions of the \dd, i.e. special \tfn s of its bases that turn one decomposition into another.

The explicit formulae for \tfn s of extremal left and right--invariant fields (\ref{rphatg},\ref{rmhatg})
and canonical variables (\ref{canplurp},\ref{canplurls},\ref{phat},\ref{canplurj}) can be used for further
investigation of particular properties of \sm s related by the \pltp y \tfn s, for example their boundary
conditions.

\section{Acknowledgements}
This work was supported by the project of the Grant Agency of the Czech Republic No. 202/06/1480 and by the
projects LC527 15397/2005--31 (L.H.) and LC06002 (L.\v S.) of the Ministry of Education of the Czech Republic.

\end{document}